\documentclass[a4paper]{jpconf}
\usepackage{graphicx}
\usepackage{pb-diagram,amsfonts,amssymb,amsthm} 
\usepackage{amsmath,epsfig}
\usepackage{cite}
\usepackage{slashed}
\def\be{\begin{equation}}
\def\ee{\end{equation}}
\def\bea{\begin{aligned}}
\def\eea{\end{aligned}}
\def\ba{\begin{eqnarray}}
\def\ea{\end{eqnarray}}

\usepackage[dvipsnames]{xcolor}

\def\yzero{\smash{\hbox{$y\kern-4pt\raise1pt\hbox{${}^\circ$}$}}}

\def\-{\hphantom{-}}

\def\s2{\frac{1}{\sqrt2}}

\def\IF{\relax{\rm I\kern-.18em F}}
\def\II{\relax{\rm I\kern-.18em I}}
\def\IP{\relax{\rm I\kern-.18em P}}
\def\IC{\relax\hbox{\kern.25em$\inbar\kern-.3em{\rm C}$}}
\def\IR{\relax{\rm I\kern-.18em R}}

\def\Dsl{\,\raise.15ex\hbox{/}\mkern-13.5mu D} 
\def\IZ{Z\kern-.4em  Z}

\begin{document}
\title{Rotating Central Charge Membranes}

\author{P. \'Alvarez$^{1,a}$, M.P. Garc\'ia del Moral$^{1,b}$, J.M. Pe\~na$^{1,c}$, R. Prado $^{1,d}$}
\address{$^1$Departamento de F\'{\i}sica, Universidad de Antofagasta, Aptdo 02800, Chile.}

\ead{$^a$pedro.alvarez@uantof.cl;$^b$maria.garciadelmoral@uantof.cl;$^c$joselen@yahoo.com
 $^d$reginaldo.prado@ua.cl} 

\begin{abstract}
In this work we obtain dynamical solutions of the bosonic sector of the supermembrane theory with central charges formulated  on $M_9\times T^2$, denoted by MIM2. The theory with this condition corresponds to a supermembrane with a $C_-$ flux. This sector of the M2-brane  is very interesting since classically is stable as it does not contain string-like spikes with zero energy and at quantum level has a purely discrete supersymmetric spectrum. We find rotating solutions of the MIM2 equations of motion fulfilling all of the constraints. By showing that the MIM2 mass operator, contains the mass operator discussed in \cite{BRR2005nonperturbative}, then we show that the rotating solutions previously found in the aforementioned work that also satisfy the topological central charge condition, are  solutions of the MIM2. Finally, we find new distinctive rotating  membrane solutions that include the presence of a new non-vanishing dynamical scalar field  defined on its worldvolume.
\end{abstract}


\section{Introduction}
In this work we obtain the first classical solutions of the equations of motion associated to the bosonic sector of the supermembrane with central charges formulated in the Light Cone Gauge (LCG) on $M_9\times T^2$. This theory has interesting properties since it has a purely discrete supersymmetric spectrum and then it can represent part of the microscopical degrees of freedom of the M-theory. In the literature there have been several studies of solutions of the membrane theory  like for example \cite{ArnlindHoppe2003,AXENIDESHoppe2017} in the context of matrix models, or like   \cite{Bozhilov2005,Bozhilov2006ExactRotatingMembraneSolutionsG2manifold,Hartnoll_2003,ARNLIND2004118SpinningMembranes} when it is formulated on a G2 or AdS backgrounds,  among others.   

Rotating membrane solutions have always interest since they can provide a preliminar signal for interpreting membranes as extended spinning particles if they become well-defined at quantum level and a proper background is chosen. They can be also be considered sources of supergravity solutions in eleven dimensions that can describe charged rotating black holes in lower dimensions, like for example \cite{CveticChong2005,Klemm2011} or they can even describe spinning solitonic solutions. Solitons have also been used in the context of rotating systems like boson stars  and Q-ball stars \cite{NontopologicalSolitonLEE1992}. 

 In \cite{nicolaihoppe} it was shown that spherical and toroidal membrane solutions could be  obtained from the membrane when it is embedded on spherical backgrounds. In \cite{FloratosPhysRevD.66.085006} new M2-brane solutions in the LCG were generalized  and in \cite{BRR2005nonperturbative} the authors found M2-brane rotating solutions on the same background $M_9\times T^2$ modelling spinning membranes. In our case we are mainly focused in the study of rotating MIM2 solutions on $M_9\times T^2$ formulated in the LCG.  We will compare our results with those of \cite{BRR2005nonperturbative} adding the topological central charge. The central charge condition on the background considered is a geometrical condition imposed on the wrapping of the membrane around the compactified target-space that induces the presence of monopoles over the worldvolume and generates a central charge in the supersymmetric algebra. The compactified supermembrane formulated  with central charges was shown in  \cite{BOULTON-2003-Discreteness} to have a purely discrete spectrum with finite multiplicity from $[0,+\infty)$. Furthermore this theory is equivalent or 'dual' to a toroidally compactified supermembrane  on a flat metric with a constant $C_{-}$ induced flux background. For all of these reasons it becomes an interesting sector to be characterized. The dynamical equations from the supermembrane theory are a system of coupled non-linear partial equations highly constrained where it is non-trivial to obtain analytical solutions. Moreover, it is known that those equations may admit have soliton solutions when they are formulated on certain backgrounds \cite{Portugues_2004}. Indeed the topological condition that we impose represents the existence of monopoles over the worldvolume of the membrane.

\section{The M2-brane with central charge}
We will consider the bosonic sector of the supermembrane with central charges formulated on  a target space $M_9\times T^2$. We will shorten it by MIM2  since it represents a supermembrane minimally immersed in the background. 
We define the embedding maps $X^m(\tau,\sigma,\rho)$ with $m=2,...,8$ on the noncompact space  $M_9$ and respectively the $X^r$, where $r=9,10$, those describing the embedding on the compactified torus. The spatial coordinates parametrizing the worldvolume $\Sigma$ are denoted by $(\sigma,\rho)$, where $\Sigma$ is a Riemann surface of genus one. The maps $X^r$ satisfy the winding condition
\begin{equation}
    \oint_{\mathcal{C}_s}dX^r=R^rm_{r}^s \,,
\end{equation}
with $m_{r}^s$ winding numbers and $R^r$ the torus radii. The MIM2 is subject to the central charge condition
\begin{equation}\label{cargaCentral}
\int_{\Sigma} d X^{r} \wedge d X^{s}=\epsilon^{r s} n A_{T^2}, \quad n \in \mathbb{Z} /\{0\}\,. 
\end{equation}
with $A_{T^2}$ represents the area of the 2-torus.
This condition implies that the one-forms associated to the embedding map of the compact sector, can be  decomposed by a Hodge decomposition $d X_r(\sigma,\rho,\tau) = d{X}_{rh}(\sigma,\rho) + dA_r(\sigma,\rho,\tau)$, with $d{X}_{rh}=R^rm_{r}^s d\widehat{X}_s(\sigma,\rho)$ a closed one form defined in terms of the harmonic forms $d\widehat{X}_s$ and $dA_r$ an exact one-form. It was  shown that  the integer $n$ associated to the central charge condition is  $n=det\mathbb{{W}}$ where $\mathbb{W}$ is the winding matrix. The central charge condition characterizes the nontrivial principle $U(1)$ bundle defined on the membrane worldvolume. The LCG Hamiltonian of the theory corresponds to \cite{GMPR2012}
\begin{equation}
\small
\begin{aligned}
\label{hamiltonianirred}
H_{MIM2}&=T^{-2/3}\int_\Sigma d^2\sigma \sqrt{W}\Big[\frac{1}{2}\Big(\frac{P_m}{\sqrt{W}}\Big)^2+\frac{1}{2}\Big(\frac{P_r}{\sqrt{W}}\Big)^2\Big]\\ 
&+ T^{-2/3}\int_\Sigma d^2\sigma \sqrt{W}\Big[ \frac{T^{2}}{4}\left\{X^m,X^n\right\}^2 + \frac{T^{2}}{2}(\mathcal{D}_r X^m)^2+\frac{T^{2}}{4}(\mathcal{F}_{rs})^2 \Big],
\end{aligned}
\end{equation}

where $\Sigma$ is a Riemann surface, $T$ is the tension of the surface of the membrane and $W$ is the determinant of the induced spatial part of the foliated metric on the membrane.  $\left\{ A,B\right\}=\frac{1}{\sqrt{W}}\epsilon^{ab}\partial_a A\partial_b B;\ $ ($ a,b=\sigma,\rho)$ is the symplectic bracket.
The canonical momentum associated to the scalar fields are $P_m$ y $P_r$. There exists a new dynamical degree of freedom associated to a one-form $\mathbb{A}=dA$ that transforms as a symplectic connection under symplectomorphisms. Associated to it, there exists a symplectic derivative and symplectic curvature  defined by
\begin{equation}
\small
\mathcal{D}_r X^m = D_r X^m +\left\{ A_r,X^m\right\}, \qquad  \mathcal{F}_{rs}= D_r A_s -D_s A_r +\left\{ A_r ,A_s \right\},  \,
\end{equation}
respectively. There, $\mathcal{D}_r$ is a symplectic connection, with $D_r\bullet=2\pi m_r^u\theta_{uv} R _r\frac{\epsilon^{\tilde{r}\tilde{s}}}{\sqrt{W}}\partial_{\tilde{r}}\widehat{X}^v\partial_{\tilde{s}}\bullet$ \,,
 $r$ index is fixed and $\theta_{uv}$ ($u,v= 9, 10$) is a matrix that has relation with the monodromy of the theory when formulated on a torus bundle \cite{GMPR2012}.
The constraints of the theory associated to the Area Preserving Diffeomorphisms  (APD) are:
\begin{equation}
\label{constraint_A}
\mathcal{D}_r \Big(\frac{P_r}{\sqrt{W}}\Big)+\left\{X^m,\frac{P_m}{\sqrt{W}}\right\}=0\,.
\end{equation}
From  (\ref{hamiltonianirred}) we derive the equations of motion for the dynamical fields $X^m(\tau,\sigma,\rho)$ and $A_r(\tau,\sigma,\rho)$
\begin{equation}
\label{equationofmotionnoncompact2}
\ddot{X}^m (\sigma, \rho, \tau)=  - \left\{\left\{X^{n}, X^{m}\right\}, X_{n}\right\}
-\left\{ \mathcal{D}^r{X}^{m}, {X}_{r}\right\}\,,
\end{equation}
\begin{equation}
\label{equationofmotionA4}
\begin{aligned}
\ddot{A}_r(\sigma, \rho, \tau) = -  \left\{X_{n}, \mathcal{D}_r X^m\right\}  -\left\{{X}_{s} ,  \mathcal{F}^{rs}\right\} \,.
\end{aligned}
\end{equation}
In order to find the admissible M2-brane solutions, the system of equations (\ref{cargaCentral}), (\ref{constraint_A}), (\ref{equationofmotionnoncompact2}) and (\ref{equationofmotionA4}) must be solved.
\section{BRR-like rotating membrane solutions from the MIM2} 

In  \cite{BRR2005nonperturbative}, the authors obtained a set of rotating membrane solutions  when a membrane is compactified on $M_9\times T^2$. They found rotating solutions associated to the following ansatz 
\begin{eqnarray}\label{AnsatzAnalogo}
X_0&=&\kappa \tau \nonumber
\\
Z_{a}(\tau , \sigma , \rho) &=& X_{2a-1}+iX_{2a}= r_{a} e^{i \beta_{a}(\tau, \sigma, \rho)}, \quad  a=1, \ldots, 4 
\\
X_{r}(\tau , \sigma , \rho) &=& R_{r}\left( n_{r} \sigma + m_{r} \rho\right) + q_{r} \tau , \ \qquad r=9.\ 10\ ,\nonumber 
\end{eqnarray}

where $r_a (\sigma,\rho,\tau)$ but we will assume to be constant, $\beta_a(\tau,\sigma,\rho)=\omega_a \tau+k_a \sigma+l_a \rho,$ with $\omega$ a rotation frequency, $k_a, l_a$ integers associated to the Fourier modes and $q_r$ integers parametrizing the Kaluza-Klein (KK) modes.

In the following, we will show that the mass operator of the MIM2 contains the BRR mass operator, and hence, contains the set of BRR solutions.
In \cite{GMPR2012} the mass operator of MIM2 was found. It has contributions from the central charge, from the KK momentum and from the Hamiltonian:
\begin{equation}\label{EnergiaMPilarr}
\mathcal{M}^2= T^2 [(2 \pi R_{11})^2 n \, (Im \tilde\tau)] ^2 
+
\left(\frac{m|q \tilde\tau-p|}{R_{11} [I_{m}\tilde\tau]}\right) ^2
+
T^{2/3}H_{MIM2} \,.
\end{equation}
In order to reproduce the BRR mass operator $\mathcal{M}_{BRR}$ from the MIM2 theory in the LCG, we assume the following ansatz:
\begin{equation}
\begin{aligned}\label{AnsatzAnalogoCFlujo}
&A_r(\tau, \sigma, \rho)= constant,\quad  
X_2(\tau, \sigma, \rho) = constant \,, \\
&Z_{a}(\tau, \sigma, \rho) = r_{a} e^{i \beta_{a}(\tau, \sigma, \rho)}, \quad \textrm{with}\quad  a=1, 2, 3;\quad 
\\
&X_{r}(\tau, \sigma, \rho) = R_{r}\left(n_{r} \sigma+m_{r} \rho\right)+q_{r} \tau + A_r(\tau, \sigma, \rho), \qquad \quad r=9, 10 \,.
\end{aligned}
\end{equation}
%
%
%
%
%
The  KK momentum contribution can be rewritten as
\begin{equation}
\left(\frac{m|q \tilde\tau-p|}{R_{11} [I_{m}\tilde\tau]}\right)=\frac{\tilde{n}_{9}^{2}}{R_{9}^{2}}+\frac{\tilde{n}_{10}^{2}}{R_{10}^{2}}=(P_9^{kk})^2+(P_{10}^{kk})^2, \qquad \text{     $\tilde{n}_{10}=m p, \ \ \tilde{n}_{9}=m q$}\,,
\end{equation}
with  $\tilde\tau=\tilde\tau_1+i\tilde\tau_2$ the Teichmuller parameter of the $T^2$ being $\tilde\tau_1=0$ and $R_{9} \equiv R_{11},\ R_{10}\equiv R_{11} Im \tilde\tau$. 
In the other hand, the central charge contribution, can be written
\begin{eqnarray}
T^2 [(2 \pi R_{11})^2 n \, (Im \tilde\tau)] ^2 &=&T_{2}^{2} 4 \pi^{2} R_{9}^{2} R_{10}^{2}\left(n_{9} m_{10}-m_{10} m _9\right)\ , 
\end{eqnarray} 
where we have used that the central charge corresponds to the determinant of the winding numbers, and the membrane tensions have been identified.

Finally, the angular momentum defined in \cite{BRR2005nonperturbative}  was identified as 
\begin{equation}
J_{a}=\int d\sigma\rho \frac{\delta S}{\delta\dot{\beta}^a}=4 \pi^{2} T_{2} r_{a}^{2} \omega_{a}\,,
\end{equation}
for the ansatz (\ref{AnsatzAnalogo}). By considering the ansatz (\ref{AnsatzAnalogoCFlujo}) and using the equation of motion associated to the $Z_a$ complexified embedding maps, to obtain the value for the frequency $\omega_c$, it can be expressed as
\begin{eqnarray}
\small
\label{eomwithansatz}
\omega_{c}^{2}-
\sum_{a=1}^{3} r_{a}^{2}  \left(k_{a} l_{c}-l_{a} k_{c}\right)^{2}-
\sum_{r=9,10} R_{r}^{2} \left(n_{r} l_{c}-m_{r} k_{c}\right)^{2}=0\,. 
\end{eqnarray}
It is then possible to reproduce the mass operator associated to the energy of the system obtained in \cite{BRR2005nonperturbative},
\begin{equation}\label{EnergiaRusso}
E= 2(4 \pi^2 T) ^{2/3} J_a\omega_a + (4 \pi^2 T) ^{2}R_9^2 R_{10}^2(n_9m_{10}-n_{10}m_9)^2+ \frac{\tilde{n}_9^2}{R_9^2}+\frac{\tilde{n}_{10}^2}{R_{9}^2} \,.
\end{equation}
Since the MIM2 is formulated in the LCG, one plane less is observed. Due to the fact that, for the ansatz considered, both of the mass operators coincide and their associated equations of motion are also reproduced.  One can realize that to reproduce \cite{BRR2005nonperturbative} results we have frozen the dynamical degree of freedom associated to the gauge symplectic connection, this also restricts the APD constraint value to the one used by \cite{BRR2005nonperturbative}. Hence those rotating solutions that also satisfy the central charge condition are also  solutions of the MIM2, whenever they satisfy (\ref{cargaCentral}).

\section{New rotating membrane solutions} 
In the following we will consider new rotating solutions to the MIM2 theory that includes the presence of a dynamical scalar field $A_r$. Due to the complexity of the equations, we consider the ansatz (\ref{AnsatzAnalogoCFlujo}) but now allowing the simplest nontrivial dependence of the gauge field, $A_r(\sigma,\tau)$. The equations of motion become reduced to:
\begin{equation}\label{EOM1}
\begin{aligned}
-\omega_{c}^{2}+\sum_{a=1}^{3} r_{a}^{2}\left(l_{c} k_{a}-k_{c} l_{a}\right)^{2}+\sum_{r=9,10} R_{r}^{2}\left(m_{r} k_{c}-n_{r} l_{c}\right)^{2}+l_{c}^{2} \sum_{r=9,10}\left(\partial_{\sigma} A_{r}\right)^{2}& \\
-2 l_{c} \sum_{r=9.10} R_{r}\left(m_{r} k_{c}-n_{r} l_{c}\right) \partial_{\sigma} A_{r} +i\ l_{c} \sum_{r=9,10} R_{r} m_{r}\left(\partial_{\sigma} \partial_{\sigma} A_{r}\right)&=0\,,
\end{aligned}
\end{equation}
for the case of $Z_c$ with $c=1,2,3$, and
\begin{equation}\label{EOM2}
\ddot{A}_{s}=-\sum_{r=9,10}\left(R_{r} m_{r}\right)^{2}\left(\partial_{\sigma} \partial_{\sigma} A_{s}\right)+\sum_{r=9,10}\left(R_{r} m_{r}\right)\left(R_{s} m_{s}\right)\left(\partial_{\sigma} \partial_{\sigma} A_{r}\right)-\sum_{a=1}^{3} r_{a}^{2} l_{a}^{2}\left(\partial_{\sigma} \partial_{\sigma} A_{s}\right)\,,
\end{equation}
for the case of $A_s$ with $s=9,10$. The APD constrain acquires this simple expression $
\sum_{r=9,10}\left(\partial_{\sigma} \dot{A}_{r} \partial_{\rho} \hat{X}^{r}\right)=0
$ and the central charge condition must also be satisfied (\ref{cargaCentral}).
We propose the following ansatz for the gauge field  $A(\sigma,\tau)=S(\sigma)+T(\tau)$.
Then, from  (\ref{EOM1}) the equation becomes
\begin{equation}
C_{1}+C_{3} S_{10}^{'}+C_{4} S^{\prime\ 2}_{10}=C_{2} S_{9}^{\prime}-C_{4} S_{9}^{\prime\ 2} \equiv \pm \lambda\,,
\end{equation}
where $S^{'} \equiv \partial_\sigma S$, and
\begin{equation}
\begin{array}{l}
C_{1}=-\omega^{2}+R_{9}^{2}\left(m_{9} k-n_{9} l\right)^{2}+R_{10}^{2}\left(m_{10} k-n_{10}l\right)^{2} \,,\quad
C_2=2 \ell R_{9}\left(m_{9} k-n_{9} \ell\right) \,, \\
C_3=2 \ell R_{10}\left(m_{10} k-n_{10} \ell\right) \,,\quad
C_4=\ell^2 \,.
\end{array}
\end{equation}

Then, depending on the value of $\lambda$, there are two possibilities:
\begin{itemize}
    \item  If $\lambda\neq0$, the solution is
 \begin{equation}
S_{9}(\sigma)= \frac{1}{l}\left(
R_{9}\left(m_{9} k-n_{9} \ell\right)
\pm
\sqrt{R_{9}^{2}\left(m_{9} k-n_{9} l\right)^{2} + \lambda}\right) \sigma + B_9 \,.
\end{equation}
 \begin{equation}
S_{10}(\sigma)=\frac{1}{l}\left(R_{10}\left(m_{10} k-n_{10} \ell\right)
\pm
\sqrt{\omega^{2} +\lambda-R_{9}^{2}\left(m_{9} k-n_{9} l\right)^{2}}\right) \sigma + B_{10} \,.
\end{equation}
Such that, when $\lambda<0$ it verify $0<\left|\lambda\right| \leq \frac{1}{2} \omega^{2}$.
    \item If $\lambda=0$ there is solution if  $\omega^{2} \geqslant R_{9}^{2}\left(m_{9} k-n_{9} l\right)^{2}$
    \begin{eqnarray}
S_9(\sigma)=0 \quad \text{or} \quad S_9(\sigma)=\frac{2}{l}R_9\left(m_{9} k-n_{9} l\right)\sigma + B_9 \,,
\end{eqnarray}
\begin{eqnarray}\small
S_{10}(\sigma)=\frac{1}{l}\left( R_{10}^{2}\left(m_{10} k-n_{10}l\right)^{2} \pm  \sqrt{\omega^{2}-R_{4}^{2}\left(m_{a} k-n_{10} \ell\right)^{2}}\right) \sigma+ B_{10} \,.
\end{eqnarray}
\end{itemize}
%
%
%
%

Next, we introduce the solution  $S_r(\sigma)=s_r\sigma+B_r$ with $s_r$, a function of $(m,n,l,k,R,\omega)$, in the equation of motion (\ref{EOM2}), this implies
\begin{equation}
\ddot{T}_r(\tau)=0,\ \Rightarrow T_r(\tau)=a_r\tau+b_r \,.   
\end{equation}
Since the  gauge field has an associated  $A_r(\tau,\sigma)$ single-valued, then we must redefine $S(\sigma)$ as  a periodic function to become well-defined. See figure 1. With these results, we can write the solution  for the gauge field as
 \begin{equation}\label{Solucionlineal}
A_r=  s_r\sigma+a_r\tau+B_r-i(2\pi s_r),\quad  \text{with} \quad \sigma \in \left[  2\pi i,2\pi (i+1)\right.) \quad i\in \mathbb{Z} \,, 
 \end{equation}
This solution is a string-like configuration and hence one can ask whether it introduces classical instabilities. In  \cite{gmr} it was shown that the MIM2 does not contain classically any flat direction at zero cost energy. Indeed it can be explicitly checked that also $(\mathcal{F}_{rs})^2\neq 0$. However since the associated one-form $dA$ is not an exact one-form since it contains infinite delta functions, hence  it does not represent an admissible symplectic gauge field nor $\mathcal{F}$ represents the symplectic curvature. If one approximate the solution by its associated truncated Fourier expansion to finite order $\widetilde{A}_r(\tau,\sigma)$, then the exactness condition is satisfied for $d\widetilde{A}_r$ and it represents a physical symplectic gauge connection with a well defined non vanishing curvature $\widetilde{F}=\epsilon^{rs} D_r\widetilde{A_s}$.    Hence, the complete solution for $Z_a$ and for approximate string-like $\widetilde{A}_r$ represents new admissible rotating solutions that all the fields of the bosonic sector of the  MIM2 theory.
\begin{figure}
     \centering
     \includegraphics[width=0.5\textwidth]{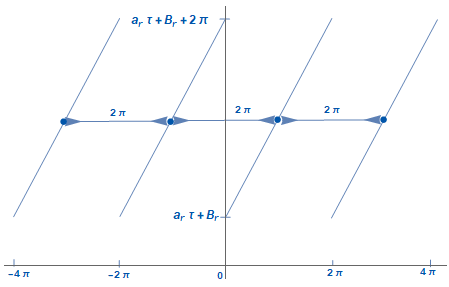}
     \caption{$A_r(\sigma,\tau)$ with $s_r=1$, and $\tau$ and $B_r$ fixed. }
     \label{fig:r}
 \end{figure}%

We have also analyzed other possibilities for the ansatz of $A$: For example if $A_r(\sigma, \tau) = a_r \sigma \tau$, $a_r =constant$, admits a solution but with the central charge equal to zero $n=0$. For the ansatz  $A_r(\sigma, \tau) = a_r f^r (\sigma) \tau^2$, the equations of motion only allow the trivial solution $f_r (\sigma)=0$. We find  (\ref{Solucionlineal}) as the only admissible string-like solution for $A_r(\sigma,\tau)$.

%

\section{Conclusions} We have characterized for first time solutions of the classical equations of  the bosonic sector of the supermembrane theory with central charges (MIM2), a sector of the M-theory with good quantum properties. We find that they admit rotating membrane solutions. The MIM2 classically does not posses string-like configurations and it is stable at quantum level. Hence the dynamics of the solutions describe an stable extended object. The solutions of the equations of motion are highly constraint because of the APD diffeomorphims and the topological condition  associated to the presence of a monopole contribution. In spite of this, we obtain that for the particular ansatz considered in 
\cite{BRR2005nonperturbative}, when it is formulated in the LCG, it is contained in the mass operator of the MIM2. Hence the rotating membrane solutions found in \cite{BRR2005nonperturbative} that also satisfy the central charge condition, are all MIM2 solutions. Furthermore we also find new rotating solutions that include the presence of a nonvanishing dynamical $A_r(\tau,\sigma,\rho)$ defined on the bosonic MIM2 worldvolume. In this work we discuss the simplest configuration that corresponds $A_r(\tau,\sigma)$. We obtain admissible string-like configurations that are linear in the spatial and time variables and satisfy all of the equations of motion and constraints. It seems that stability arguments of the solutions and the central charge condition avoids that more general solutions like $A(\sigma,\tau)=\alpha f(\sigma) g(\tau)$ to be allowed. In order to define a single-valued function a periodicity is imposed that renders it as a kind of shawtooth function whose associated one-form contains infinite delta functions. It is possible to define an approximate string-like solution $\widetilde{A}$ defined in terms of its Fourier expansion truncated  to finite order. The associated one-form  $d\widetilde{A}$ corresponds to a well-defined symplectic gauge field with a non- vanishing curvature  $\widetilde{\mathcal{F}}^2\ne 0$. This work is part of larger study in which we also intend to characterize further the existence of new solutions of the membrane with fluxes \cite{M2Rotating}.
\section{{Acknowledgements}}  The authors are very grateful to A. Restuccia for helpful discussions.
R. P. y J.M.P. thank to the projects ANT1956 y ANT1955 of the U. Antofagasta. All the authors want to thank to SEM18-02 project of the U. Antofagasta. MPGM, R. Prado  thank to the international ICTP project NT08 for kind support.

\section*{References}

\bibliography{References}

\end{document}